\documentstyle[12pt]{article}
\normalbaselines
\begin{document}
\vbox {\vspace{6mm}}

\begin{center}

{\large \bf POLARIZATION STRUCTURE\\
OF QUANTUM LIGHT FIELDS: A NEW INSIGHT.\\
2: GENERALIZED COHERENT STATES, SQUEEZING AND GEOMETRIC PHASES} \\[7mm]

                   V.P.KARASSIOV\\
{\it Lebedev Physical Institute, Leninsky prospect 53, Moscow\\
117924, RUSSIA  Internet: karas@sci.fian.msk.su}\\[5mm]
\end{center}
\vspace{2mm}
\begin{abstract}

Within the new description of the polarization structure of quantum light
(given in Part I) some types of generalized coherent states related to
the  polarization $SU(2)_p$ group are examined. With their help we give a
quasiclassical description of polarization properties of light fields
and discuss the concept of squeezing and uncertainty relations for
multimode  light in the polarization quantum optics. As a consequence,
a new classification of polarization  states of quantum
light is  obtained. We also derive geometric phases acquired by different
quantum light beams transmitted through "polarization rotators".

\end{abstract}
\vbox{\vspace{10mm}}


\vbox{\vspace{10mm}}

Short title: POLARIZATION STRUCTURE OF QUANTUM LIGHT-2

\vbox{\vspace{10mm}}


\vbox{\vspace{15mm}}

 Submitted to Journal of Physics, ser. A on March 07 1995
\newpage
\section{ Introduction}

Recently a new formalism[1-3] was proposed for a description of polarization
structure of multimode quantum light fields using the polarization $SU(2)_p$
symmetry and a related concept of the $P$-quasispin which generalizes
the Stokes vector notion at the quantum level and is closely related to the
Stokes operators defined in [4]. This approach enabled us to gain a new
insight into the polarization structure of light and quantum mechanisms of
its depolarization (see Part I (ref. [2]) and references therein).

At the same time, so-called squeezed states of light are intensively examined
now within quantum optics (see, e.g., [5-10] and references therein) since
these states have attractive properties of the "noise reduction" in
measurements of some quantum mechanical observables. However, we note that
squeezed states have been studied sufficiently only for single-mode fields
[5-7] whereas for multimode fields it is not the case since even the
definition of the concept of multimode squeezing is not unique due to a
variety of the choices of measurable quantities and appropriate uncertainty
measures for them[9-11].

 It is well known [11] that the generalized coherent states (GCS)
 $$|\{\zeta_i\};\psi_0\rangle=D(g(\{\zeta_i\})) |\psi_0\rangle, \eqno (1.1)$$
generated by the action of the displacement operators $D(g(\{\zeta_i\}))=
\exp (\sum_i \zeta_i F_i)$ of the groups $G^{DS}$ on certain fixed reference
vectors $|\psi_0\rangle$ in the given spaces $L^D$ of the representations
$D = \{ D(g), g\in G^{DS} \}$ of  the groups $G^{DS}$, present an effective
tool for the study of quantum systems having the dynamic symmetry groups
$G^{DS}$. In particular, the average values $\langle\{\alpha_i\};\psi_0|
f(\{ F_i\})|\{\alpha_i\};\psi_0\rangle$ of the arbitrary functions
$f(\{ F_i\})$, corresponding to the observables and depending on the
generators $F_i$ of $G^{DS}$, as well as the quasiprobability distribution
functions ($Q$-functions)
$$Q(\{\alpha_i\};\psi_0;\rho)=
\langle\{\alpha_i\};\psi_0|\rho|\{\alpha_i\};\psi_0\rangle, \eqno (1.2)$$
with $\rho$ being the density matrix, are widely used for vizualizing
squeezed states and for a description of quasiclassical properties of the
appropriate quantum systems near the ``classical limit'' [3,6,10-14].
For example, in quantum optics similar quantities, defined using the
familiar Glauber's CS and associated with Weyl-Heisenberg group $W(m)$, are
widely used for the description of $m$-mode electromagnetic fields [6,11,13].
The GCS associated with $SU(m)$ groups play the same role for the systems of
$n$-level emitters of radiation [11,15]. Furthermore, the GCS formalism
appeared to be useful to calculate geometric phases of quantum systems
evolving in time on manifolds with a nontrivial topology [16-19].

  The aim of this paper is to examine different kinds of GCS associated with
the $SU(2)_p$ group and to apply them for a quasiclassical description of
polarization properties of quantum light beams, an analysis of the concept of
squeezing of the multimode light related to polarization degrees of freedom
and for calculations of specific geometric phases acquired by quantum light
beams transmitted through "polarization rotators". In Section 2, for the sake
of self-consistency of the exposition, some main points of Part I and related
papers are  recapitulated. In Section 3 we define and study some sets of the
$SU(2)_p$ GCS and determine the $Q$-representations of different polarization
operators providing a quasiclassical description of polarization in quantum
optics. In Section 4 the problem of squeezing in polarization quantum optics
is discussed using the abovementioned GCS. In particular, we show that
quantum states of light beams generated by specific unpolarized biphoton
clusters of the $X$-type (see Part I) exhibit, in a sense, an absolute
squeezing in polarization degrees of freedom. As a consequence, a new
classification of polarization states of light within quantum optics is
obtained. In Section 5 we calculate "polarization" geometric phases acquired
by quantum light beams in different pure states after their transmission
through "polarization rotators". In Section 6 some prospects of further
developments and applications of results obtained are briefly discussed.

\section{ Preliminaries}

 As it was shown in [1,2], in polarization quantum optics there are specific
observables which characterize proper polarization properties of light
beams and correspond  to the group $U(2)$ of a specific polarization gauge
invariance of light fields. The generators of this group $U(2)$ in the
helicity $(\pm)$ polarization basis are of the form
$$  P_0 = { 1\over {2}} \sum _{i=1}^{m} [ a_+^+(i)a_+(i)-a_-^+(i) a_-(i)] =
\sum _i  P_0(i),
\quad P_\pm = \sum _{i=1}^{m} a_{\pm}^+((i)a_{\pm}(i)=\sum _i  P_{\pm}(i),$$
$$ 2P_2=i(P_+ -  P_-),\quad2P_1 =(P_+ + P_-),\quad N=
\sum _{i=1}^{m} \sum_{\alpha = +,-}a_{\alpha} ^+(i)a_{\alpha}(i) =
\sum_{i}N(i)
 \eqno      (2.1)$$
where $m$ is the number of spatiotemporal (ST) modes under study, $N$ is
the total photon number operator and operators $P_{\alpha }$ are
generators of the $SU(2)_p$ subgroup defining the polarization $(P)$
(quasi)spin. The operators $P_{\beta }$ and $N$ satisfy commutation relations
$$ [ N,P_{\alpha} ] = 0,\quad [P_0,P_{\pm}]=\pm P_{\pm},\quad [P_+,P_-]=2P_0
          \eqno      (2.2)$$
and in the case $m=1$ coincide up to the factor $1/2$ with Stokes operators
$\Sigma_{\alpha }:\Sigma_1 =2P_2,  \Sigma_2 = -2P_0,\Sigma_3 = -2P_1$ [4].
As is clear from Eqs. (2.1) the total $P$-quasispin components of the  field
is the sum of the appropriate quasispin quantities for single ST modes.
However, from the experimental viewpoint the total $P$-quasispin enables us
to examine new interesting physical phenomena connected with correlations of
different modes, in particular, "entangled states" which are widely discussed
in multiparticle interferometry [20-22]. Components $P_{\alpha}$ of the
$P$-quasispin are parameterized on the so-called Poincar\'e sphere $S_{P}^2$
in the classical statistical optics [3,19] and are measurable in
polarization experiments related to counting photons with definite circular
($P_0$) or linear ($P_i, i=1,2$) polarizations in quantum optics [23].

Quantities $<P_{\alpha }>, <N>$ determine the polarization degree $degP$ of
light beams with arbitrary wave fronts and frequencies by the relation [2,10]
$$degP =2 [\sum_{\alpha =0,1,2}(<P_{\alpha}>)^{2}]^{1/2}/<N>   \eqno (2.3)$$
generalizing the appropriate definition [4,13] for the case of a single ST
mode. At the same time the quantum averages $<|P^2|>=\bar{p}(\bar{p}+1)$ of
the $SU(2)_{p}$  Casimir operator $P^2 = (1/2)(P_+P_- + P_-P_+) + P^2_0$
are connected by the relations
$$ a) E^2_r =P^2 -P_0(P_0 -1), \eqno (2.4a)$$
$$b)<|P^2|>= \bar{p}(\bar{p}+1)=
\sum _{\alpha =0,1,2}[\sigma^P _{\alpha}+(<|P_{\alpha}|>)^{2}]=
\sum _{\alpha =0,1,2}\sigma^P _{\alpha} +[degP<N>/2]^2
\eqno  (2.4b)$$
with the so-called "radial" operator $E_r =\sqrt{P_+P_-}$ used for examining
phase properties  of electromagnetic fields [24] and with the variances
$$\sigma^P _{\alpha } = <|P^2_{\alpha }|> - (<|P_{\alpha }|>)^2
\eqno (2.4c)$$
determining different uncertainty measures for operators $P_{\alpha}$
("polarization noises") [2,10-12].

The $2m$-mode Fock space
 $$L_F(2m)=Span\{|\{n_{\pm}(i)\}> =
\prod_{j=1}^m(a_+^+(j))^{n^+_j}(a^+_-(j))^{n^-_j}|0\rangle\}$$
is  decomposed in a direct sum
$$ L_F(2m)=\sum_{p,\sigma}L^{(p,\sigma)}=\sum _{p,n,\lambda}L(p,n,\lambda)
       \eqno     (2.5)$$
of the $SU(2)_p$-invariant  subspaces $L(p,n,\lambda)$ which are specified
by eigenvalues $p,n,\lambda =[\lambda_i]$ of the $P$-spin, $N$ and a set of
operators describing non-polarization degrees of freedom and are spanned by
basis vectors $|p \mu;n, \lambda>$ which are eigenvectors of the operators
$P^2, P_0, N$ [25]:
$$ P^2|p\mu;n,\lambda >=p(p+1)|p\mu;n,\lambda>, \quad P_0 |p\mu;n,\lambda>=
\mu|p\mu;n,\lambda>,$$
$$ \ N \mid p\mu;n, \lambda> = n \mid p\mu;n, \lambda>  \eqno (2.6)$$
The vectors $|p, \mu;n, \lambda>$ may be expressed in the form of polynomials
in operators $a_{\pm }^+(i), Y_{ij}^{+}, X^{+}_{ij}$ acting on the vacuum
vector$|0>$ where operators
$$ Y^{+}_{ij}={1\over {2}}(a_+^+(i)a_-^+(j)+a_-^+(i)a_+^+(j)),
X^{+}_{ij}=a_+^+(i)a_-^+(j)-a_-^+(i)a_+^+(j) \eqno          (2.7)$$
are  solutions of the operator equations
$$[P_0,Y^{+}_{ij}]=0;\quad [P_{\alpha },X^{+}_{ij}]=0,\quad \alpha=0,+,-
                                                         \eqno       (2.8)$$
and may be interpreted as creation operators of $P_0$-scalar and $P$-scalar
biphoton kinematic clusters respectively. For example, in the cases $m=1,2$
we have the following expressions [1,3]
$$a)|p\mu> =[(p-\mu)!(p+\mu)!]^{-1/2}(a^{+}_{+}(1))^{|\mu|+\mu}
(a^{+}_{-}(1))^{|\mu|-\mu} (Y_{11}^{+})^{p-|\mu|}|0>, \quad  n=2p,
   \eqno (2.9a)$$
$$b)|p,\mu;n, \lambda =t> =$$
$$[\frac{(2p+1)(p+\mu)!(p-\mu)!(p-t)!(p+t)!}{(n/2+p+1)!(n/2-p)!}]^{1/2}$$
$$\sum_{\alpha}\frac{ (a_{+}^{+}(1))^{p+\mu-\alpha}
(a_{-}^{+}(1))^{t-\mu+\alpha}(a_{+}^{+}(2))^{\alpha}
(a_{-}^{+}(2))^{p-t-\alpha}}{(\alpha)!(p-t-\alpha)!
(p+\mu-\alpha)!(t+\alpha-\mu)!}(X^{+}_{12})^{n/2-p}|0>
 \eqno  (2.9b)$$
where $2t =n(1)-n(2)=n_+(1)+n_-(1)-n_+(2)-n_-(2)$ is the difference
of the photon numbers in the first and second ST modes. In general,
the states $|p,\mu;n,\lambda >$, whose explicit forms (in terms of the
$SU(2)$ generating invariants) can be found in [10,15,25], describe
light beams representing a mixture of both usual photons and $P$- and
$P_0$-scalar biphotons[1,2].

Biphoton operators $X_{ij}, X^+_{ij}$  generate the Lie algebra
$so^*(2m)$  commuting with the polarization invariance algebra
$su(2)_{p}=Span\{P_{\alpha}\}$[2]. Therefore, states $|\psi >$,
belonging to a subspace $L(p\mu )$ of states with given $p, \mu $ at
initial time, will evolve in this subspace $L(p\mu)$ under action of the
interaction Hamiltonians
$$H_{X} = H^{'}_{int}(\{ X_{ij},X^+_{ij}; E_{ij}\}) \eqno (2.10a)$$
describing some anisotropic parametric processes. Extending the algebra
$so^*(2m)$ by adding operators $Y_{ij}, Y^+_{ij}$ we get the algebra
$u(m,m)$ commuting with the polariztion subalgebra $u(1)=Span\{P_0\}$ and
associated with interaction Hamiltonians
$$H_{X,Y}=H^"_{int}(\{ Y_{ij}, Y^+_{ij}; X_{ij}, X^+_{ij}; E_{ij}\})
\eqno (2.10b)$$
which keep invariant for time evolution subspaces $L^{'}(\mu) =
\sum_{p\geq \|\mu|} L(p\mu)$ [10]. The algebra $u(m,m)$ contains the
subalgebra $sp(2m,R)$ generated by biphoton operators $Y_{ij}, Y^+_{ij}$.

\section{Generalized coherent states and quasiclassical description of light
polarization}

 In this section, developing results [1,3,19], we examine in the $2m$-mode
Fock space $L_F(2m)$ different types of polarization GCS associated with
the $SU(2)_p$ group orbits and useful in applications.

As is well known, general GCS of the $SU(2)$ group orbit type are defined
in accordance with (1.1) as follows [11]
$$
|\xi ;\psi_0\rangle \equiv |\theta,\varphi;\psi_0\rangle= D(g(\xi(\vec{n})))
|\psi_0\rangle, D(g(\xi))\equiv \exp (\xi J_+ -\xi^* J_-)
\eqno (3.1) $$
where $\xi(\vec{n})=-\theta /2 \exp(-i\varphi), \quad 0\le \theta \le\pi,
\quad 0\le \varphi \le 2\pi$ are the angular coordinates of the unit vector
$\vec{n} =(\sin \theta \cos \varphi, \sin \theta \sin \varphi, \cos \theta)$
determining a position of the``classical'' quasispin $\vec J=(J_\alpha)$ on
its Poincar\'e sphere $S^2_{P}(\theta,\varphi)$; $|\psi_0\rangle$ is a
certain reference vector in the space $L$ of the states of the system.
{}From the physical viewpoint, the states $|\xi ;\psi_0\rangle$ describe
output light beams obtained by means of action of quantum
"$SU(2)$-rotators" with Hamiltonians
$$H_{SU(2)}= g J_+ +g^* J_-     \eqno (3.2)$$
on the input beams in the quantum state $|\psi_0\rangle$ (see, e.g.,
[18,26,27] and references therein for possible realizations of such rotators
in experimental devices).

For spin systems having a fixed spin value $j$ one of the basis vectors
$|jm\rangle$ of the irreducible representation (irrep) $D^j(SU(2))$
is used as $|\psi_0\rangle$, and the values of $m=\pm j$ correspond to the
GCS most near to classical states [11]. A peculiarity of the polarization
quasispin $(J_\alpha = P_\alpha)$ of the light fields is that according to
Eq. (2.5) the Fock spaces $L_F(2m)$ may be viewed as direct sums of the
specific $SU(2)$ fiber bundles and contain the subspaces $L^{j\sigma}$ of
the irrep $D^j(SU(2))$ with $j=p=0,1/2,1,$\dots , generally (for $m \geq 2$)
with a certain multiplicity $\sigma $. Hence, in order to get the
``complete polarization portrait`` of the quantum light field
and to calculate different physical quantities defined on the whole
space $L_F(2m)$ one should have complete sets of the  $SU(2)_p$ GCS
(3.1) with a set of reference vectors
$|\psi_0\rangle=|\psi_0^{p,\sigma}\rangle\in L^{p,\sigma},
\quad p=0,1/2,\dots$, or with a set of vectors
$|\psi_0\rangle=|\psi_{0,\gamma}\rangle$, having nonzero projections on
each of the subspaces $L^{P,\sigma},\quad p=0,1/2,...$ that, e.g., occurs
for states describing physical light beams. Below we consider some examples
of both types GCS in both mathematical and physical aspects.

In the first case, using the decomposition (2.5), it seems natural to choose
for $|\psi_0^{p,\sigma}\rangle$ the vectors $|p\mu;n,\lambda \rangle$ from
Eq. (2.6). Then, making use of the definition (3.1) and of the transformation
properties of the operators $a^+_{\pm}(j), Y^+_{ij}, X^+_{ij}$ with respect
to the $SU(2)_p$ group transformations $D(g(\xi))$ from (3.1) [1,10],
$$a) a^+_{\pm}(j)\longrightarrow \tilde{a}^+_{\pm}(j)
\equiv(\eta^\pm(\theta,\varphi)a^+(j))
= a^+_\pm(j)\cos {\theta\over 2} \pm a^+_\mp(j) \exp(\pm i\varphi)
\sin{\theta\over 2}, \eqno (3.3a)$$
$$b) Y^+_{ij} \longrightarrow \tilde{Y}^+_{ij} =
Y^+_{ij} \cos {\theta} +1/2 \sin {\theta} [ a^+_+(i) a^+_+(j)\exp(i\varphi) +
a^+_-(i) a^+_-(j) \exp(-i\varphi)],  \eqno (3.3b)$$
$$c) X^+_{ij} \longrightarrow X^+_{ij}
\eqno (3.3c)$$
we get the sets $\{|\theta,\varphi;p,\mu,n,\lambda \rangle\}$ of the
polarization GCS generated by the reference vectors $|p,\mu;n,\lambda
\rangle$ and reproducing their form in terms of "$SU(2)$-rotated"
operators (3.3) [1,3]; herewith the operator $(\eta^\pm(\theta,\varphi),
a^+(j))$ may be interpreted as the creation operator of the elliptically
polarized photon in the $j$-th ST mode having the ellipticity parameters
determined by the angles $\theta,\varphi$ [28]. Evidently, using the
definition (3.1) and Eq. (3.3) one can also obtain expansions of GCS
$\{|\theta,\varphi;p,\mu,n,\lambda \rangle\}$ in terms of initial states
$|p,\mu;n,\lambda \rangle$:
$$|\theta,\varphi;p,\mu,n,\lambda \rangle
=\sum_{\mu '}u^p_{\mu '\mu}(\theta,\varphi)|p\mu';n,\lambda \rangle, $$
$$u^p_{\mu '\mu}(\theta,\varphi)=[\frac{(p+\mu)!(p-\mu')!}
{(p+\mu')!(p-\mu)!}]^{1/2} \frac{\exp(i\varphi(\mu-\mu'))}{(\mu-\mu')!}
(\tan\frac{\theta}{2})^{\mu-\mu'}(\cos \frac{\theta}{2})^{2p}$$
$$F(\mu-p,-p-\mu';\mu-\mu'+1; -\tan^2\frac{\theta}{2})
  \eqno (3.4)$$
where the expansion coefficients
$u^p_{\mu '\mu}(\theta,\varphi)$ are particular $SU(2)\; D$-functions and
$F(a,b;c; z)$ is the Gauss hypergeometric function.

The states $\{|\theta,\varphi;p,\mu,n,\lambda \rangle\}$ belong , from
the mathematical point of view, to the class of {\it semi-coherent} ones
(which are coherent (quasiclassical) in polarization degrees of freedom
and orthonormalized (strongly quantum) in other ones) [1] since their
overlap integral, easily calculated with the help of Eq. (3.4), has the form
$$
\langle\theta,\varphi;p,\mu, n,\lambda|\theta',\varphi';
p',\mu', n',\lambda '\rangle \equiv I_{p, p';n, n';\mu,\mu';\lambda,\lambda'}
(\eta^{\pm}, \eta'^{\pm})= $$
$$\delta_{p p'}\delta_{n n'}\delta_{\lambda \lambda '}\;
\sum_{\mu"}[u^p_{\mu"\mu}(\theta,\varphi)]^* u^p_{\mu"\mu'}
(\theta',\varphi')=\delta_{p p'}\delta_{n n'}\delta_{\lambda \lambda '}\;
[\frac{(p-\mu)!(p+\mu')!}{(p+\mu)!(p-\mu')!}]^{1/2}\times$$
$$ \frac{(\eta'^{+} \eta ^{+*})^{p+\mu}(\eta'^{-} \eta ^{-*})^{p-\mu'}
(\eta'^{+} \eta ^{-*})^{\mu'-\mu}}{(\mu'-\mu)!}
F (-p-\mu', -p+\mu;1+\mu'-\mu; -|\frac{(\eta'^{+} \eta^{-*})}
{(\eta'^{+} \eta^{+*})}|^2)       \eqno (3.5)$$
where $F(a,b;c; z)$ is the Gauss hypergeometric function and
$$(\eta'^{+} \eta ^{+*})=\cos\frac{\theta}{2}\cos\frac{\theta'}{2}
+\sin\frac{\theta}{2}\sin\frac{\theta'}{2}\exp( i(\varphi'- \varphi)=
(\eta'^{-} \eta ^{-*})^*,$$
$$(\eta'^{-} \eta ^{+*})=-\exp(-i\varphi)'\cos\frac{\theta}{2}\sin
\frac{\theta'}{2}+\sin\frac{\theta}{2}\cos\frac{\theta'}{2}
\exp(- \varphi)=-(\eta'^{+} \eta ^{-*})^*    \eqno (3.5')$$

The GCS sets $\{|\theta,\varphi;p,\mu,n,\lambda \rangle\}$ contain subsets
of states
$$|\theta,\varphi;p,n,\lambda \rangle_\pm \equiv
\exp(\xi P_+-\xi^*P_-)|p,\pm p;n,\lambda\rangle = $$
$$ \sum_{\mu}[\frac{(2p)!}{(p+\mu)!(p-\mu)!}]^{1/2}
 (\pm \sin\frac{\theta}{2})^{p\mp\mu}(\cos\frac{\theta}{2})^{p\pm\mu}
 \exp[-i(\mu\mp p)\varphi] |p,\mu;n,\lambda \rangle
 \eqno (3.6) $$
which satisfy the maximal classicality criterion [11] in the polarization
degrees of freedom and are (over)complete in $L_F(2m)$ yielding the
following decomposition of the identity operator $\hat I$ [3,15]:
$$
\hat I = \sum_{n,p,\lambda} \int_0^\pi \int_0^{2\pi}{(2p+1)
\over 4\pi} \sin \theta d\theta d\varphi
|\theta,\varphi;p,n,\lambda\rangle_\pm\langle\theta,\varphi;p,n,\lambda|_\pm
\eqno (3.7)$$
that provides possibilites of calculating different physical averages on
$L_F(2m)$ using these GCS. We also note that GCS (3.6) in their form are
specific generating functions for states $|p,\mu;n,\lambda \rangle$ [15,25];
therefore, using states (3.6) as reference vectors in Eq. (3.1) (with
another parameter $\xi(\vec{n'})=-\theta'/2 \exp(-i\varphi')$) we obtain
(after the substitution $\theta=\pi /2, \exp(-i\varphi)=z$) generating
functions for  GCS (3.4) that can be used in concrete calculations. Moreover,
both sets $\{|\theta,\varphi;p,n,\lambda \rangle_\pm\}$ are equivalent  from
the mathematical viewpoint as it is seen from the formal equality
$$|\theta,\varphi;p,n,\lambda \rangle_{-} = \exp(-i2p\varphi)
|\theta +\pi,\varphi;p,n,\lambda \rangle_{+}    \eqno (3.6') $$
which follows from Eq. (3.6). Therefore, hereafter we will use, as a rule,
only the set $\{|\theta +\pi,\varphi;p,n,\lambda \rangle_{+}\}$
omitting for the sake of simplicity the subscript $"+"$.

The construction of Eq. (3.6) is simplified in the picture of
independent ST modes, when the group $SU(2)_p$ acts in the space $L_F(2)$ of
each $j$-th ST mode independently, and its action is determined by the
angles $(\theta_j,\varphi_j)$ characterzing by "partial" $P$-quasispin
components $P_{\alpha}(j)$ and Poincar\'e spheres $S^2_{P}(j)$:
$$
|\{\theta_j,\varphi_j\};\{n_j\}\rangle_\pm  \equiv \prod^m_{j=1}
\exp(\xi_jP_+(j)-\xi^*_jP_-(j))(a^+_{\pm}(j))^{n_j}[n_j!]^{-1/2}|0\rangle
 =  \prod_{j=1}^m
{(\eta^\pm(\theta_j,\varphi_j),a^+(j))^{n_j}\over [ n_j!]^{1/2}}|0\rangle
\eqno (3.8)$$
The set of GCS (3.8) is complete (an analog of Eq.~(3.7) is valid for it)
and yields the ``polarization phase portrait of the field'' adequate to
independent measurements for each ST mode. The connection between the
sets (3.6) and (3.8) is realized via the generalized Clebsh-Gordan
coefficients of $SU(2)_p$ [15]. Note that choosing arbitrary Fock states
$ \prod^m_{j=1} (a_+^+(j))^{n^+_j}(a^+_-(j))^{n^-_j}|0\rangle$ as the
reference vector in (3.1) we obtain polarization GCS
$$
|\{\theta,\varphi\};\{n^+_j;n^-_j\}\rangle =
 \prod_{j=1}^m{(\eta^-(\theta,\varphi),a^+(j))^{n^-_j}
(\eta^+(\theta,\varphi),a^+(j))^{n^+_j}\over [n^+_j!n^-_j!]^{1/2}}
|0\rangle  \eqno (3.8^*)$$
which are similar to states (3.8) in their form and may be used for
expanding other kinds of polarization GCS in series using appropriate
expansions of the reference vectors in terms of the Fock states.
The GCS (3.8)  are the Fock states in terms of the "rotated" photon
operators $(\eta^\pm(\theta,\varphi),a^+(j))$ and are unitarily equivalent
to to the initial Fock states; hence there are some difficulties to
produce them (as well as states $|\theta,\varphi;p,\mu,n,\lambda \rangle$)
in physical experiments[13]. Therefore, from the physical viewpoint it
is of interest  to consider other types of GCS of $SU(2)_p$, which do not
contain the discrete parameters $n,\lambda$ labeling $SU(2)$-invariant
subspaces $L^{(p,\sigma={n,\lambda}}$.  So, if taking in (3.1) the
reference vectors $|\psi_0^{p,\sigma}\rangle$ of the form[1]
$$|\psi_0^{p,\{\zeta_{ij},\kappa_{ij} \}}\rangle_{\pm}=$$
$$\exp(\sum_{i<j}[\kappa_{ij}E_{ij}+ \zeta_{ij}X^+_{ij}-\kappa^*_{ij}E_{ji}-
\zeta^*_{ij}X_{ij}])
(a^+_\pm(1))^{2p}[(2p)!]^{-1/2}|0\rangle, E_{ij}\equiv \sum_{\alpha =\pm}
a^+_{\alpha}(i)a_{\alpha}(j)  \eqno (3.9)$$
we obtain states
$$|p;\{\zeta_{ij},\kappa_{ij} \};\theta,\varphi\rangle_{\pm}=$$
$$\exp(\sum_{i<j}[\kappa_{ij}E_{ij}+ \zeta_{ij}X^+_{ij}-\kappa^*_{ij}E_{ji}-
\zeta^*_{ij}X_{ij}])
(\eta^\pm(\theta,\varphi),a^+_\pm(1))^{2p}[(2p)!]^{-1/2}|0\rangle
\eqno (3.10)$$
which are  GCS with respect to both the polarization $SU(2)_p$ and the
"biphoton" $SO^*(2m)$ groups, (over)complete in $L_F(2m)$, provide a
quasiclassical description of both polarization and biphoton degrees of
freedom[10] and are generated in processes governed by Hamiltonians (2.10a)
and (3.2). Note that using the "disentangling theorems"[11,29] for the
$SO^*(2m)$ displacement operators $\exp(\sum_{i<j}[\kappa_{ij}E_{ij}+
\zeta_{ij}X^+_{ij}-\kappa^*_{ij}E_{ji}-\zeta^*_{ij}X_{ij}])$ one can obtain
expansions of states (3.10) in series of states (3.6). For example,
in the simplest non-trivial case $m=2$, when $SO^*(4)=SU(2)\otimes
SU(1,1),\;SU(2)=Span\{J_+=E_{12}, J_-=E_{21},J_0=1/2(E_{11}-E_{22})\},
SU(1,1)= Span\{K_+=X^+_{12},K_-= X_{12}, K_0+=1/2(E_{11}+E_{22})+1\}$,
with the help of results [11,29,15] we find [1]
$$|p;\{\zeta, \kappa\};\theta,\varphi\rangle_{\pm}= \exp([\kappa E_{12} -
\kappa^* E_{21}]) \exp([\zeta X^+_{12}-\zeta^* X_{12}])
(\eta^\pm(\theta,\varphi),a^+_\pm(1))^{2p}[(2p)!]^{-1/2}|0\rangle=$$
$$[\cosh |\zeta|]^{-2(p+1)}[\cos |\kappa|]^{2p} \sum_{T,\tau}
(\tanh|\zeta|\exp(i\arg\zeta))^T(-\tan|\kappa|\exp(-i\arg\kappa))^{\tau}$$
$$[\frac {(T+2p+1)!}{(2p+1)(T)!(2p-\tau)! \tau!}]^{1/2}
|\theta,\varphi;p,n=2(T+p),t=p-\tau \rangle_\pm \eqno (3.11)$$
{}From the physical point of view  it is also of interest to consider
generalizations of GCS (3.10) related to Hamiltonians (2.10b) and (3.1)
and obtained when  replacing the reference vector (3.9) by the $U(m,m)$
group GCS
$$|\psi_0^{p,\{\zeta_{ij},\kappa_{ij}, \gamma_{ij} \}}\rangle_{\pm}=$$
$$\exp(\sum_{i<j}[\kappa_{ij}E_{ij}+ \zeta_{ij}X^+_{ij}+\gamma_{ij}Y^+_{ij}
-\kappa^*_{ij}E_{ji}-\zeta^*_{ij}X_{ij}-\gamma^*_{ij}Y_{ij}])
(a^+_\pm(1))^{2p}[(2p)!]^{-1/2}|0\rangle,
 \eqno (3.12)$$
Without dwelling on a detailed analysis of such GCS we write
down their expressions (cf. (2.9a))
$$
|p;\gamma;\theta,\varphi\rangle_\pm =\exp(\gamma \tilde{Y}^{+}_{11}-\gamma^*
\tilde{Y}_{11})\frac{(\tilde{a}^+_{\pm})^{2p}}{\sqrt(2p)!}|0\rangle=$$
$$[\cosh |\gamma|]^{-(2p+1)} \sum_{\tau}
(\tanh|\gamma|\exp(i\arg\gamma))^{\tau}
[\frac {(2p+\tau)!}{(2p)! \tau!}]^{1/2}
|\theta,\varphi;p,\mu= \pm p,n=2(\tau+p) \rangle
\eqno (3.13)$$
for the states which are GCS of $SU(2)_p\otimes
SU(1,1),\;SU(1,1)= Span\{K'_+=Y^+_{11},K'_-= Y_{11}, K'_0+=1/2(E_{11}+1)\}$
and describe in the case of $p=0$ the twin-photon beams of
unpolarized light obtained in degenerate parametric processes [2,10,23].

An alternate type of "physical" polarization GCS may be obtained if
one takes in (3.1) the sets of reference vectors
$|\psi_{0,\gamma}\rangle$, having the nonzero projections on all
$L^{p,\sigma}$. A natural example of such a set is the familiar set of
Glauber's coherent states:
$$
|\{\alpha^+_j,\alpha^-_j\}\rangle=\prod^m_{j=1} \exp
[\alpha^+_ja^+_+(j)+\alpha^-_ja^+_-(j)-
(\alpha^+_j)^*a_+(j)-(\alpha^-_j)^*a_-(j)]|0\rangle.
\eqno (3.14)$$
Then using the definition (3.1) and the $SU(2)_p$ transformation properties
(3.3) of $a^+_\pm(j)$, one gets from Eq. (3.14) the set of GCS
$$
|\theta,\varphi;\{\alpha_j^+,\alpha_j^-\}\rangle
  \equiv  \exp(\xi P_+-\xi^* P_-)
|\{\alpha^+_j,\alpha^-_j\}\rangle=|\{\tilde \alpha^+_j(\theta,\varphi),
\tilde \alpha^-_j(\theta,\varphi)\}\rangle,$$
$$
\tilde \alpha^\pm_j(\theta,\varphi)  =  \alpha^\pm_j\cos\frac{\theta}{2}
\mp\exp(\mp i\varphi)\alpha^\mp_j\sin\frac{\theta}{2}, \eqno (3.15)
$$
which can be obtained expirementally by action of quantum polarization
"rotators" on the initial states (3.14).

The states (3.15) are analogous to the initial set (3.14), but with two extra
(redundant from the mathematical viewpoint) parameters involved, namely,
$\theta$ and $\varphi$. This redundance can be removed by imposing two
constraints on parameters $\alpha^{\pm}_j$ in (3.14). For example, in the
case of $m=1$ one may choose the subsets of (3.15) in the form ($j$ is
fixed):
$$
|\theta_j,\varphi_j;\alpha^+_j\rangle_+\equiv
|\theta_j,\varphi_j;\alpha^+_j,0\rangle = |\alpha^+_j\cos\frac{\theta_j}{2},
\alpha^+_j\exp(i\varphi_j)\sin\frac{\theta_j}{2}\rangle,\quad
|\theta_j,\varphi_j;\alpha^-_j\rangle_-\equiv
|\theta_j,\varphi_j;0,\alpha^-_j\rangle,
\eqno (3.16)$$
which describe  the elliptically polarized waves and coincide with usual
Glauber CS (3.14) for $m=1$ but with picking out polarization
coordinates $\theta_j,\varphi_j$ explicitly (unlike the form (3.14)) that
it is important, e.g., for constructing polarization $Q$-functions [3,19].
 In the general  case $m\ge2$ the subsets (3.16) are not complete in
$L_F(2m)$ that, however, is unimportant from the physical point of view.
(The complete sets of GCS of such a type in $L_F(2m)$ may be obtained, e.g.,
by taking  $m$-fold product of GCS (3.16) [3,19].)
Another possibility is related to subsets  of GCS (3.15) where
states (3.14) are constrained by conditions
$$
\sum_{j=1}^m arg\alpha^+_j = 0, \quad \sum_{j=1}^m arg\alpha^-_j = 0
\eqno   (3.17)$$
which determine for $m=1$ an alternate to (3.16) set  of polarization
GCS. Note that from the physical point of view one can also  determine
other types of "physical" polarization GCS obtained via actions of
"polarization rotators" on different physical input states , e.g.,
eigenstates of the biphoton destruction operators $Y_{ij}, X_{ij}$, etc.
[1,30] that, however, is beyond the scope of the paper.

The sets of GCS obtained above may be  used for the quasiclassical
analysis of the polarization properties of quantum light fields. In
particular, one can use the definition (1.2) to introduce the complete
polarization $Q$-functions [3] as follows
$$
Q(\theta,\varphi;\psi_0;\rho)\equiv
\mbox{Tr}[\rho |\theta,\varphi;\psi_0\rangle\langle\theta,\varphi;\psi_0|]=
\langle\theta,\varphi;\psi_0|\rho|\theta,\varphi;\psi_0\rangle,
\eqno  (3.18) $$
where $\rho$ is the complete density operator for the state of the
field, $|\theta,\varphi;\psi_0\rangle$ being defined by Eq. (3.1). Then,
substituting the specifications (3.6), (3.8), (3.10), (3.12), (3.15)
and  (3.16) for $|\theta,\varphi;\psi_0\rangle$ into Eq.~(3.18), we get
the appropriate concrete  types of the complete polarization
quasiprobability functions. Note, however,  that such functions,
besides the dependence on the polarization parameters $\theta,
\varphi$, involve the additional quantum numbers $n,\lambda,
\{\alpha^\pm_j\}$, etc., which characterize the non-polarization
degrees of freedom of the field. Therefore, to get its ``pure polarization
quasiclassical portrait'' in the $\rho$-state it is sufficiently to make use
of the reduced polarization quasiprobability functions $Q^p (\varphi,\theta;
\psi_0;\rho)$, resulting from Eq.~(3.18) after the summation (or integration)
over non-polarization variables. Such functions determine "error bodies" and
may be used to analyse the ``polarization squeezing'' [10] in analogy with
the familiar $Q$-functions in the case of the standard quadrature squeezing
[6]. Keeping in mind the completeness of GCS (3.6), one may determine for
this aim only one type of $Q^p$-functions based on GCS (3.6):
$$
Q^{p}(\theta,\varphi;p;\rho)= \sum_{n,\lambda}\langle\theta,\varphi;
p,n,\lambda |\rho|\theta,\varphi;p,n,\lambda \rangle,
\eqno  (3.18^{*}) $$

Let us  calculate some of such $Q^p$-functions, substituting in $(3.18^{*})$
concrete  density operators $\rho_i=|\theta',\varphi';\psi_0^i\rangle
\langle\theta',\varphi';\psi_0^i| $ for pure states
$|\theta',\varphi';\psi_0^i\rangle, i=1,2,$ described by
Eqs. (3.4) and (3.15) (and restricting oneself for the sake of simplicity
by the case of  $m=2$ in (3.15)). Then, after some algebra one gets the
following expressions for the appropriate $Q^p$-functions
$$
a)Q^{p}(\theta,\varphi;p;\rho_1)=  \delta_{p p'}
\;[\frac{(2p)!}{(p+\mu')!(p-\mu')!}]|(\eta'^{+} \eta ^{+*})|^{2(p+ \mu')}
|(\eta'^{-} \eta ^{+*})|^{2(p- \mu')}=$$
$$ [\cos^2\frac{\theta-\theta'}{2}-\sin^2\frac{\varphi'-\varphi}{2}
\sin{\theta}\sin{\theta'}]^{p+ \mu'}[\sin^2\frac{\theta-\theta'}{2}
+\sin^2\frac{\varphi'-\varphi}{2}
\sin{\theta}\sin{\theta'}]^{p- \mu'},
\eqno  (3.19a) $$
 $$b)Q^{p}(\theta,\varphi;p;\rho_2) =  \frac{\exp(-\sum_{i=1,2}
[|\alpha_i^+|^2 +|\alpha_i^-|^2]) \;(2p+1)J_{1+2p}(-2|[\alpha_1 \alpha_2]|)}
{|[\alpha_1 \alpha_2]|^{2p+1} }\times$$
$$[(\sum_{i=1,2}|\alpha_i^+|^2)|(\eta'^{+} \eta ^{+*})|^2 +
(\sum_{i=1,2}|\alpha_i^-|^2)|(\eta'^{-} \eta ^{+*})|^2
+ 2 Re[(\sum_{i=1,2}\alpha_i^+\alpha_i^{-*})(\eta'^{+} \eta^{+*})
(\eta'^{-} \eta^{+*})^*]]^{2p}, $$
$$ [\alpha_1 \alpha_2]=\alpha_1^+ \alpha_2^-
-\alpha_1^- \alpha_2^+        \eqno  (3.19b) $$
where $J_{1+2p}(-2|[\alpha_1 \alpha_2]|)$ is the Bessel function.
Note that is due to the structure of Eqs (3.11), (3.13)  angular
dependences of $Q^{p}(\theta,\varphi;p;\rho_i), i=3,4,$  for
states described by these equations can be easily obtained  from (3.19a).
 For comparison we also write down  the $Q^p$-function
$$
e)Q^{p}(\theta,\varphi;p;\rho_{th}(1))= [1- \exp(-\beta)]^2
\exp(-2p\beta),$$
$$\rho_{th}(1)=[1- \exp(-\beta)]^2 \sum_{n,\mu} \exp(-n\beta)
|p=n/2 \mu> <p=n/2 \mu|,\quad  \beta = (kT)^{-1}
\eqno  (3.20) $$
for the case of a single ST mode in the mixed state $\rho_{th}(1)$ [13] of
the thermal equilibrium.

Similarly, one can determine $Q^p$- representations
$$f(\{P_{\alpha}\};\theta,\varphi;\psi_0)\equiv
\langle\theta,\varphi;\psi_0|f(\{P_{\alpha}\})|
\theta,\varphi;\psi_0\rangle \eqno (3.21)$$
 for arbitrary polarization operators $f(\{P_{\alpha}\})$ using for this aim
 polarization characteristic functions [11]
 $$\chi^{\psi_0}_{\{P_{\alpha}\}}(\{\nu_i\})= \langle\theta,\varphi;\psi_0|
 \prod_i \exp(\nu_i P_i)|\theta,\varphi;\psi_0\rangle \eqno (3.22)$$
where exponents are taken in an order. Calculations of the (3.22) right
sides, evidently, are reduced to finding overlap integrals like Eq. (3.5).
Without dwelling on a detailed discussion  of this question we find
some characteristic functions which are useful
in applications. In particular, in studies of squeezing problems it is
necessary to have $Q$- representations for lowest powers of $P_{\alpha}$
(see,e.g., [5-10] and the following Section). Keeping  also in mind
expansions of the type (3.11), (3.13) it is sufficiently for this aim
to calculate particular characteristic functions
$\chi^{\psi_0}_{P_{\alpha}}(\mu_{\alpha})$ (with $\alpha$ being fixed) for
GCS (3.4)  and (3.15) (or (3.14)). Then, using the definitions
$2P_1=(P_+ + P_-), 2P_2 = i(P_+ -  P_-)$  and Eqs. (3.3)- (3.5),
(3.15) one finds
$$a)\chi^{sc_{p,\mu}}_{P_{\alpha =1,2}}(\nu_{k=1,2})\equiv
\langle\theta,\varphi;p,\mu; n,\lambda| \exp(\nu_k P_k)
|\theta,\varphi;p,\mu; n,\lambda \rangle=$$
$$\sum_{\alpha}\frac{(p+\mu)!(p-\mu)!}{(p+\mu-\alpha)!(p-\mu-\alpha)!\alpha!
\alpha!}[\sin^{2}\frac{\tau}{2} (\sin^{2}\theta \sin^{2}(\frac{\pi k}{2}-
\varphi) -1)]^{\alpha}\times$$
$$[\cos\frac{\tau}{2}+i\sin\frac{\tau}{2}\sin\theta \sin(\frac{\pi k}{2}-
\varphi)]^{p+\mu-\alpha}[\cos\frac{\tau}{2}-i\sin\frac{\tau}{2}\sin\theta
\sin(\frac{\pi k}{2}-\varphi)]^{p-\mu-\alpha},
                             \eqno (3.23a)$$
$$b)\chi^{Gcs}_{P_{\alpha =1,2}}(\nu_{k=1,2})\equiv
\langle\{ \alpha^+_j(\theta,\varphi), \alpha^-_j(\theta,\varphi)\}|
  \exp(\nu_k P_k)|\{ \alpha^+_j(\theta,\varphi),
 \alpha^-_j(\theta,\varphi)\}\rangle =$$
 $$\exp \left[(\cos\frac{\tau}{2}-1)\sum_{j=1}^m
\{|\alpha_j^+(\theta,\varphi)|^2
+|\alpha_j^-(\theta,\varphi)|^2 \} \right]$$
$$\times\exp\left[i\sin\frac{\tau}{2}\sum_{j=1}^m
 2Im\{\alpha_j^-(\theta,\varphi)
(\alpha_j^+(\theta,\varphi))^* \exp(\frac{-i\pi k}{2})\}\right],
\eqno (3.23b)$$
where $\nu_k=i\tau, \;\tau$ is purely real and $k=1 (2)$ for $P_{1 (2)}$.
Similarly, using a diagonal analog of the transformations (3.3), one gets
$$a)\chi^{sc_{p,\mu}}_{P_0}(\nu_0=i\tau)=\sum_{\alpha}\frac{(p+\mu)!(p-\mu)!}
{(p+\mu-\alpha)!(p-\mu-\alpha)!\alpha!\alpha!} [-\sin^{2}\frac{\tau}{2}
\sin^{2}\theta]^{\alpha}\times$$
$$[e^{\frac{-i\tau}{2}}\sin^2\frac{\theta}{2}+e^{\frac{i\tau}{2}}
\cos^2\frac{\theta}{2}]^{p+\mu-\alpha}
[e{\frac{i\tau}{2}}\sin^2\frac{\theta}{2}+e^{\frac{-i\tau}{2}}
\cos^2\frac{\theta}{2}]^{p-\mu-\alpha},   \eqno (3.24a)$$
$$b)\chi^{Gcs}_{P_0}(\nu_0=i\tau)= $$
 $$\exp \left[-\sum_{j=1}^m\{|\alpha_j^+(\theta,\varphi)|^2
+|\alpha_j^-(\theta,\varphi)|^2 \} \right]
\exp\left[\sum_{j=1}^m\{|\alpha_j^+(\theta,\varphi)|^2\exp(i\tau)
+|\alpha_j^-(\theta,\varphi)|^2 \exp(-i\tau)\} \right], \eqno (3.24b)$$

\section{Squeezing and a new classification of polarization states of light
in quantum optics}

The decomposition (2.5) and the results obtained in the previous Section
yield an effective tool for studies of the squeezing problems of multimode
light beams with consideration of polarization that, in turn, implies a new
classification of the polarization states of quantum light fields [10].
In fact, a definition of squeezing in quantum mechanics is based on an
analysis of different uncertainty relations for a set $\{A_i, i=1,...,r>1\}$
of non-commuting Hermitian operators $A_i$ representing some quantum
observables [5-10,31-34]. These relations are associated with specific
measures of admissible quantum fluctuations ("noises") for observables
$A_{i}$ in the state $|>$ expressed in terms of expectations
$<|(A_{i})^{s}|>$ characterizing differences between quantum observables
$A_{i}$ and their classical analogs[11,12,31-33]. For example, the most
widespread uncertainty relation (of the Weyl-Heisenberg type) has the
form [11]
$$
\Delta A_{i} \Delta A_{j} \geq 1/2 |<|[A_{i}, A_{j}]|>|  \eqno (4.1)
$$
where $(\Delta A)^2\equiv\sigma_{A}=<|(A)^2|>-(<|A|>)^2$ is a standard
quadratic measure (variance) of a deviation of the quantum quantity
$A$ from its classical analog ($<|A|>$). Specifically, for the single-mode
electromagnetic field one makes use of two quadrature components of the field
$A_1=(a^++a)/\sqrt{2}, A_2=i(a^+-a)/\sqrt{2}$ as observables $A_{i}$, and
$|<|[A_1, A_2]|>|=1$ determines a boundary (vacuum or zero-point) level of
admissible quantum field fluctuations [5]. Then conditions {\it a) $S^A_{12}
\equiv \Delta A_{1}\Delta A_{2}\longrightarrow min$ with constraints (4.1)
(a joint quasiclassical behaviour of $A_1$ and $A_2$)} and {\it b)
$\Delta A_1$ (or $\Delta A_2$) $< [(S^A_{12})_{min}]^{1/2}$ (a suppression
of one quadrature noise)} define the usual one-mode field quadrature
squeezing realized with the help of GCS $\exp(za^{+2}-z^{*}a^2)|\psi>$ of
the group $SU(1,1)\sim Sp(2,R) = Lin \{ L_0 =a^+a/2 +1/4, L_+ =(a^+)^2/2,
L_- =a^2/2\}$ conserving the canonical commutation relation $[a,a^+]=1$ and
(when $z$ is real) the "error area" $S^A_{12}$ [5-7]. (Emphasize an
importance of the condition a) in the definition above and of the $S^A_{12}$
invariance with respect to $\exp(z[a^{+2}-a^2])$ ($z$ is real) for a search
of squeezed states; a relaxation of the first requirement leads to a "soft"
(with a fixed value of $S^A_{12}$) squeezing notion whereas a rejection
of both ones contradicts principal original ideas [5,6] and makes the
class of squeezed states too large.)

However, for multimode fields the situation becomes more complicated as
in this case we have a more vast set of observables which obey non-trivial
commutation relations, and there  exist many possibilities of definition of
squeezing related to different choices (from physical considerations) of some
subsets of observables, adequate joint uncertainty measures for them and some
boundary (or reference) levels of admissible quantum fluctuations [8-10,34].
Specifically, in polarization quantum optics as such subsets, besides
different field quadrature components [8,9], one may also take components of
the $P$-quasispin obeying the commutation relations of the $su(2)_p$ algebra
and subsets of unpolarized biphoton operators of $X$- and $Y$- types
generating the $so^*(2m)\subset u(m,m)\supset sp(2m,R)$ algebras associated
with Hamiltonians (2.10). That enables to define (when maintaining basic
features of the concept above) different sorts of multimode light squeezing
related to appropriate degrees of freedom. Without dwelling on all aspects
of this vast topic we focus here our attention on features (including
definitions) of specific kinds of squeezing related to polarization (and
partially to biphoton) degrees of freedom applying for this purpose standard
uncertainty measures and their general analysis for arbitrary Lie algebras
[11-12] as well as the GCS techniques of the algebras above.

First of all we note that the conditions (4.1) are less restrictive for
generators $A_i$ of arbitrary Lie algebras $g$ than for one-mode field
quadrature components since, in general, right sides of these inequalities
are not fixed $c$-numbers but depend on quantum states under consideration.
Furthermore, adequate measures of a joint quasiclassical behaviour of
the set $\{A_i\}$ are the $g$-invariant quantities $(\Delta A)^2=\sum g_{ij}
[<A_iA_j>-<A_i><A_j>]$ ($g_{ij}$ is the Killing-Cartan metric tensor)
related to Casimir operators of $g$ rather than  $S^A_{ij}$ [11,12].
Therefore, in this case the squeezing definition above has to be
modified (when retaining its basic features), e.g., as {\it a
suppression of one or more "partial noises" $\Delta A_i$ at a minimal or
a given reference level of the "collective noise" $(\Delta A)^2$
consistent with conditions (4.1)}. A natural search of appropriate
squeezed states may be realized in a set of the $g$ GCS conserving
values $(\Delta A)^2$ and the structure relations of $g$ that provides
(together with the modified definition) a group-theoretical treatment
of the squeezing concept for $\{A_i\}\subset g$.

For example, a "purely polarization" squeezing is defined in such a manner by
means of a minimization of a $SU(2)$-invariant "radial" uncertainty measure
$(\Delta P)^2\equiv \sum_{\alpha}\sigma^P _{\alpha}= \bar{p}(\bar{p}+1)-
[degP <|N|>/2]^2$ of total polarization noises or its normalized version
$(\delta P)^2 = (\Delta P)^2 /(<|N|>)^2$ (determinig a level of polarization
quasiclassicality of the field in a given pure quantum state $|>$) together
with an analysis of the  relations (4.1) for $A_i =P_i, i=1,2,0$. For fixed
values $p$ of the polarization quasispin it may be realized on GCS (3.6) (or
on their linear combinations over discrete parameters $n, \lambda$) [10].
Furthermore, owing to the complementarity (commutativity of actions) of the
algebras $so^*(2m)$ and $su(2)_p$ on $L_F(2m)$ (cf. (2.8)) this definition
is completely compatible with the "$X$-biphoton" squeezing defined similarly
(but with peculiarities due to a $(\Delta X)^2$ definition) for observables
of the $so^*(2m)$ algebra, and, according to the analysis [11,12], GCS (3.10)
realize such a joint "polarization - $X$-biphoton" squeezing; therefore,
operators $S_P(\{\xi\})\equiv\exp(\xi P_+-\xi^*P_-)$ and $S_X(\zeta_{ij}\})
\equiv \exp(\sum_{i<j}[\zeta_{ij}X^+_{ij}-\zeta^*_{ij}X_{ij}])$ may be
called, respectively, as polarization and $X$-biphoton squeezed operators.
An extra peculiarity follows in polarization squeezing studies from the
fact that in accordance with Eq. (2.5) physical states describing light
beams do not belong to a single irreducible subspace of $su(2)_p$;
besides, in polarization optics there exist only two different basic
measurement procedures related to linear ($P_1$ or $P_2$) and circular
($P_0$) polarization types. Bearing in mind these general remarks we examine
below different polarization GCS of the previous Section as test
functions (with $p, \theta, \varphi$ being variables) to determine
different (related to possible specifications of a suppression of
"partial noises" $\Delta P_i$) types of polarization squeezed states.

By analogy with the usual one-mode quadrature squeezing we consider
at first  polarization noises  for the  Glauber's GCS (3.15). Using the
characteristic functions (3.23b) and (3.24b) one finds relations
$$a) (\Delta P_{1})^2 = (\Delta P_{2})^2 = (\Delta P_{0})^2 =
1/4\sum_{j=1}^m\{|\alpha_j^+|^2+|\alpha_j^-|^2 \}=<|N|>/4,$$
$$(\Delta P)^2 = 3<|N|>/4, \quad (\delta P)^2 = \frac{3}{4<|N|>}
\eqno  (4.2a)$$
$$
b)|<|P_{0}|>|=1/2|\sum_{j=1}^m\{[|\alpha_j^+|^2-|\alpha_j^-|^2]\cos\theta -
2Re[\alpha_j^-\alpha_j^{+*}\exp(-i\varphi)]\sin\theta\}|,$$
$$|<|P_{2}|>|=$$
$$1/2|\sum_{j=1}^m \{[|\alpha_j^+|^2-|\alpha_j^-|^2]\sin\theta \sin\varphi
+2Im[\alpha_j^-\alpha_j^{+*}]\cos^2\frac{\theta}{2}+
2Im[\alpha_j^-\alpha_j^{+*}e^{-i2\varphi}]\sin^2\frac{\theta}{2}\}|,$$
$$|<|P_{1}|>| =$$
$$1/2|\sum_{j=1}^m \{[|\alpha_j^+|^2-|\alpha_j^-|^2]\sin\theta \cos\varphi +
 2 Re[\alpha_j^-\alpha_j^{+*}]\cos^2\frac{\theta}{2}+
 2 Re[\alpha_j^-\alpha_j^{+*}e^{-i2\varphi}]\sin^2\frac{\theta}{2}\}|
                      \eqno (4.2b)$$
manifesting an absence of any polarization squeezing because all partial
polarization noises $(\Delta P_{\alpha})^2$ are equal to a quarter of the
standard (Poissonian) quantum level $<|N|>$ of noises in optics [5-7] for
any values of both polarization ($\varphi, \theta$) and coherent
($\alpha_j^{\pm}$) parameters; this feature distinguishes actions of
polarization squeezed operators $S_P(\{\xi\})$ on GCS (3.14) from those for
usual quadrature squeezed operators $\exp(za^{+2}-z^{*}a^2)$ (cf. [5-7]).

However, the situation is quite different for GCS (3.6), (3.10). Indeed,
according to Eqs. (3.23a) and (3.24a), these GCS with a fixed value $p$
minimize the uncertainty measure $(\Delta P)^2$:
$$a) (\Delta P)^2 =(\Delta P)^2 _{min} =p \; \leq \;<N>/2, \quad
(\delta P)^2 \leq \;\frac{1}{2<N>}     \eqno  (4.3a)$$
and  relations (4.1) for $A_i =P_i, i=1,2,0$ have on these GCS the form:
$$
b)\Delta P_{1} \Delta P_{2} = p/2\; [\cos^2\theta +\frac{1}{4}\sin^4\theta
\sin^2 2\varphi]^{1/2}\geq 1/2|<|P_{0}|>| = p/2\; |\cos\theta|, $$
$$
\Delta P_{1}\Delta P_{0}=p/2\;|\sin\theta|[1-(\sin\theta\cos\varphi)^2]^{1/2}
\geq 1/2 |<|P_{2}|>| =p/2\; |\sin\theta\sin\varphi|,$$
$$
\Delta P_{2}\Delta P_{0}=p/2\; |\sin\theta|[1-(\sin\theta\sin\varphi)^2]^{1/2}
\geq 1/2 |<|P_{2}|>| =p/2\; |\sin\theta\cos\varphi|     \eqno (4.3b)$$
As is seen from Eqs. (4.2b), for $\theta=0,\pi$ and any $\varphi$ we get a
complete suppression of the circular polarization noise  $\Delta P_{0}:
\Delta P_{0} =0$  whereas $(\Delta P_{1})^2 =(\Delta P_{2})^2 =p/2$;
therefore, GCS (3.6), (3.10) for $\theta=0,\pi$ manifest a circular
polarization squeezing unlike the circular coherent states (3.15)
characterized by the condition: $\sum_{j=1}^m \alpha_j^-\alpha_j^{+*} =0$.
Similarly, GCS (3.6), (3.10) for $\theta=\pi/2, \varphi=0 \mbox{or} \pi/2$
manifest a linear polarization squeezing. Another ("circular-linear") type
of polarization squeezing may be determined by imposing the constraints:
$$ (\Delta P_{a})^2 \leq \; (\Delta P)^2/3 = p/3 \leq\; <N>/6,
\quad  a =0,1
                      \eqno (4.4a)$$
that corresponds to a joint suppression of one linear ($\Delta P_{1}$) and
the circular ($\Delta P_{0}$) partial noises at the expense of increasing
$\Delta P_{2}$. As it follows from Eqs. (4.3), (4.4a) such a situation is
attained for parameters $\varphi, \theta$ constrained by the conditions
$$ \frac{1}{3} \leq \; \sin^2 \theta \cos^2 \varphi\leq \; \frac{4}{9},
\qquad \frac{1}{2} \leq \; \sin^2 \theta \leq \; \frac{2}{3},
\quad \frac{1}{2} \leq \; \cos^2 \varphi \leq \; 1       \eqno (4.4b)$$

However, in practice, it is more or less easy to produce only GCS (3.10)
with $p=0$ which provide an absolute polarization squeezing for any
$\theta, \varphi$ and describe an absolutely unpolarized light
("polarization vacuum") characterized by relations [2,10]
$$<|(P_{\alpha })^s|>= 0,\quad \alpha = 0,1,2,\; s\geq 1  \eqno  (4.5a) $$
showing the full absence of appropriate polarization noises
($<|(P_{\alpha })^s|>-(<|(P_{\alpha })|>)^s$) of any order measured by
appropriate noises of difference photocurrents in schemes [23]. (For
comparison we point out that for usual quantum spin systems characterized
by a fixed value $j$ of angular momentum such an absolute squeezing for
the $su(2)$ algebra observables is realized on the only vector $|j=0,
m=0>$.) Note that for the case $m=2$ Eqs. (4.5a) imply relations
$$<|(P_{\alpha}(1))^s|>= (-1)^s <|(P_{\alpha}(2))^s|>, $$
$$<|(P_{\alpha}(1))^s|><|(P_{\beta}(1))^s|> = <|(P_{\alpha}(2))^s|>
<|(P_{\beta}(2))^s|>          \eqno          (4.5b)$$
that it is of interest for studies of the EPR-paradox, entangled states
and "hidden variable" theories [20-22]. Besides, the polarization vacuum
is the only sort of quantum light having the property $(\Delta P)^2 = 0$
as it follows from Eqs. (2.4b) and (4.1).
The situation is somewhat different for GCS (3.12) also realized with
the help of parametric oscillator generators (but corresponding to the
Hamiltonians (2.10b)). In fact, as the $u(m,m)\supset sp(2m,R)$ algebras
commute only with the $u(1) =Span\{P_0\}$ subalgebra of $su(2)_p$, the
polarization ($S_P(\{\xi\})$) and $Y$-biphoton ($S_Y(\{\gamma_{ij}\})\equiv
\exp [\sum (\gamma_{ij}Y^{+}_{ij} -\gamma ^*_{ij}Y_{ij})]$) squeezed
operators act in general not independently (as it follows from Eq. (3.3)).
Therefore, for the particular case of GCS (3.13) generated by biphotons
of the $Y$ -type Eqs. (4.3) are replaced by relations
$$a)(\Delta P)^2 =p +\frac{1}{2}(p+1)(2p+1)\sinh^2|2\gamma|=
\frac{(1-2p)[<N>(p+1) + p] +<N>^2/2}{2p+1},   \eqno (4.6a)$$
$$
b)\Delta P_{1} \Delta P_{2} =\frac{1}{2}(\Delta P)^2
[\cos^2\theta +\frac{1}{4}\sin^4\theta\sin^2 2\varphi]^{1/2}
\geq1/2|<|P_{0}|>|= 1/2p|\cos\theta|,$$
$$\Delta P_{1} \Delta P_{0} = \frac{1}{2}(\Delta P)^2
[1-\sin^2\theta\cos^2 \varphi]^{1/2} |\sin\theta|
\geq 1/2 |<|P_{2}|>|= 1/2p|\sin\theta\sin \varphi|,$$
$$\Delta P_{2} \Delta P_{0} =  \frac{1}{2}(\Delta P)^2
[1-\sin^2\theta\sin^2 \varphi]^{1/2} |\sin\theta|
\geq  1/2 |<|P_{1}|>|=1/2p|\sin\theta\cos\varphi|  \eqno (4.6b)$$
depending on both polarization ($\varphi, \theta$) and biphoton ($\gamma$)
parameters. Evidently, repeating main steps of the Eqs. (4.3) analysis,
one may determine "soft" (because of differences between Eq. (4.6a) and
Eq. (4.3a)) versions of different types of polarization squeezing
considered above. Specifically, for $\gamma \not=0$ (determined, e.g.,
by conditions of squeezing biphoton or quadrature variables) Eq. (4.6a) is
minimized when $p=0$ on states describing another (in comparison with GCS
(3.10)) class of unpolarized quantum light (so-called twin-photon light with
hidden polarization [26]) for which
$$(\Delta P)^2 = \frac{1}{2}\sinh^2|2\gamma|=<N> +<N>^2/2, \quad
(\delta P)^2 = \frac{1}{<N>} + 1/2    \eqno (4.6a^*)     $$
and Eqs. (4.5a) are valid only either for $s=1$, any $\alpha$
or for $\alpha =0, \theta =0,\pi$, any $s, \varphi$ [2,10].

So, we have shown that polarization GCS generated by means of actions of
squeezed operators $S_P(\{\xi\})\; S_X(\zeta_{ij}\})$ and $S_P(\{\xi\})\;
S_Y(\{\gamma_{ij}\})$ on extremal vectors of the $SU(2)_p$ irreps with
fixed values of $p$ determine different classes of polarization squeezed
states characterized by values of the total polarization noises
$(\Delta P)^2$ (determining a "hardness" of squeezing) and of partial
coefficients $k^P_i=(\Delta P_i)^2 /(\Delta P)^2$ (indicators of polarization
squeezing in a given class of GCS). Furthermore, maximal suppressions of the
total polarization noises are attained for states corresponding to quantum
unpolarized light. However, actions of operators $S_P(\{\xi\})$ on
Glauber's CS (3.14) (with $\alpha_{i}^{\pm} \neq 0$), including states of
coherent unpolarized light characterized by special values of parameters
$\alpha_{i}^{\pm}$:
$$\sum_{j=1}^m \{[|\alpha_j^+|^2-|\alpha_j^-|^2] = 0, \quad
\sum_{j=1}^m [\alpha_j^-\alpha_j^{+*}] =0, \eqno   (4.7)$$
do not lead to GCS displaying any polarization squeezing, although GCS (3.15)
have a bigger degree of "polarization quasiclassicality" in comparison with
GCS (3.12)-(3.13) (cf. Eqs. (4.2a) and (4.6a)).

This situation is similar to that for thermal unpolarized light as is seen
from  counterparts of Eqs. (4.2), (4.7) for states with $\rho_{th}(1)$
of Eq. (3.20) [23]:
$$a) (\Delta P)^2 = \frac{3}{2}\exp(-\beta)[1-\exp(-\beta)]^{-2}
= \frac{3}{4}(<N>+ <N>^2/2),$$
$$ (\Delta P_{\alpha})^2 = \frac{1}{4}(<N>+ <N>^2/2), \; \alpha = 1,2,0
\eqno  (4.8a)$$
$$b) |<|P_{\gamma}|>|=0 , \; \gamma= 0,1,2 \eqno (4.8b)$$
At the same time, a phase randomization of GCS (3.14) satisfying the
condition $|\alpha_{i}^{+}|=|\alpha_{i}^{-}|$, leads to an intermediate type
of unpolarized light revealing features of a "soft" squeezing (but due to
another mechanism in comparison with quantum correlations for GCS (3.10)-
(3.13)). Specifically, in the simplest case of one ST mode, described by
the density matrix
$$ \rho_{c/r}= (2\pi)^{-1}\int d\chi
|\{\alpha^{\pm}_1\}><\{\alpha^{\pm}_1\}|, \alpha^+_1 =\alpha^-_1
exp(i\chi),         \eqno(4.9)$$
analogs of Eqs. (4.2), (4.8) take the form [23]
$$a) (\Delta P)^2 = \frac{1}{4}(3<N>+ <N>^2), \quad
(\Delta P_{0})^2 = \frac{1}{4}<N>, $$
$$ (\Delta P_{\alpha})^2 = \frac{1}{4}(<N>+ <N>^2/2), \; \alpha = 1,2,
\eqno  (4.10a)$$
$$b) |<|P_{\gamma}|>|=0 , \; \gamma= 0,1,2 \eqno (4.10b)$$
All this leads to a new classification of states of unpolarized light within
quantum optics[10]; herewith different classes of unpolarized light are
distinguished by an availability (or not) of polarization (and biphoton)
squeezing and its "hardness" as well as by values of the quantitative
characteristics of light depolarization $dep_P =(1- 2\bar{P}/\bar{N})$ and
$dep_{P_{0}}= (1-|2\bar{\pi}|/\bar{N})$ yielded by the $P$-spin formalism
[2,10].

A similar (but characterized additionally by values of the polarization
degree $degP$) classification can be obtained for partially
polarized light as it follows from the analysis above for GCS with $p\not=0$.
However, in real physical experimental situations states of
light beams do not belong to a single subspace $L(p \mu)$ but are
superpositions of states from different subspaces $L(p \mu)$. Therefore,
it is of interest to study polarization squeezing properties of partially
polarized light beams obtained by actions of the biphoton squeezed operators
$S_Y, S_X$ together with the polarization squeezed operators $S_{P} (\xi)$
on states $|in>$ of some input light beams. As a result we can obtain new
classes of non-classical states of partially polarized light. Without
dwelling here on an analysis of this topic we note that due to Eq. (2.8)
actions of operators $S_X$ do not change $(\Delta P_{\alpha})^2$ but decrease
$(\delta P)^2$ in comparison with those of input beams; however, it is not
the case for actions of operators $S_Y$.

\section{Geometric phases of polarizaton coherent states}

Another area of applications of results obtained in Section 3 is
calculations of Pancharatnam-type geometric phases acquired
by  polarization GCS during the cyclic evolution on the Poincar\'e sphere.
In classical polarization optics it is well known [25,35], that
during the cyclic evolution of its polarization state the
classical plane wave acquires an additional phase shift equal to
half the solid angle subtended by the trajectory of the tip of the
Stokes vector on the Poincar\'e sphere. This additional phase is
shown to be a particular case of the Pancharatnam's phase [36]
associated with the $SU(2)$ symmetry of the polarization states.
It is invariant with respect to deformations of the trajectory
leaving the solid angle unchanged and, therefore, is of purely
geometric nature. Pancharatnam's ideas have been used [18] to set a
generalized definition of the geometric phase, valid for a
wide class of quantum evolutions, generally, non-cyclic.

 A natural question arises, what happens to the states of quantum light
in the similar situation. The considerations presented above make it
possible to apply the general definitions [18] of the geometric phase
$\gamma$ to the polarization GCS, since the angles $\theta, \varphi$
enter the appropriate expressions explicitly as classical parameters[19]:
$$
\gamma=-\oint_C A_sds, \eqno (5.1)$$
where the gauge potential $A_s$ is expressed as
$$
A_s  =  Im\langle\xi(\theta,\varphi),\psi_0 |\frac{d}{ds}
|\xi(\theta,\varphi),\psi_0\rangle \nonumber  =
Im\langle\xi(\theta,\varphi),\psi_0|\nabla_
{\vec\Omega}|\xi(\theta,\varphi),\psi_0\rangle{d\vec \Omega\over ds},
\eqno (5.2)$$
$s$ is an evolution variable which determines the motion of the system
along the evolution trajectory $C$. The states $|\xi(\theta,\varphi),
\psi_0\rangle$ are supposed to be the normalized polarization GCS defined
by Eq.~(3.1) with a certain particular choice of the reference state
vectors mentioned above. Then, using explicit forms of the derivatives on
the unit sphere,
$$a)
{d\vec \Omega\over ds}=
\vec e_\theta\frac{d\theta}{ds}+\vec e_\varphi
\sin\theta\frac{d\varphi}{ds},\eqno (5.3a)$$
$$b)
\langle\xi(\theta,\varphi),\psi_0|\nabla_{\vec\Omega}|\xi(\theta,
\varphi),\psi_0\rangle=
[\vec {e}_\theta\frac{\partial}{\partial\theta' }+
\frac{\vec e_\varphi}{\sin\theta}
\frac{\partial}{\partial\varphi' }]
\langle\xi(\theta,\varphi),\psi_0|\xi(\theta',\varphi'),\psi_0\rangle|
_{\theta'= \theta, \varphi'=\varphi},
\eqno (5.3b)$$
and expressions for GCS overlap integrals one can calculate appropriaite
geometric phases (5.1).

As the first example let us consider the general semi-coherent polarization
GCS $|\theta,\varphi;p,\mu, n,\lambda\rangle$. Then, substituting the
expression (3.5) for the overlap integral in Eq.~(5.3b), calculating
derivatives  and using the definition (5.1)-(5.2) we find
$$
\gamma= -2\mu \oint_C\sin^2\frac{\theta}{2}d\varphi, \eqno (5.4)$$
which is the $-2\mu$ multiple to half the solid angle subtended by $C$ on
the Poincar\'e sphere. In particular, for $\mu=1/2$ this result coincides
with that for  classical plane waves, and  for $\mu=0$ we get the
complete absence of any geometric phase. A similar result can be found for
other polarization GCS associated with the decomposition (2.5) of the Fock
space. For example, for the GCS (3.10), (3.13) one gets
$$
\gamma= \mp 2p \oint_C\sin^2\frac{\theta}{2}d\varphi, \eqno (5.5)$$
where  $ p = n/2$ for $m=1$ and $0\leq p \leq n/2$ for $m\geq2$ whereas for
GCS $(3.8^*)$ the factor $\mp 2p$ is replaced by $\sum_j (n^+_j - n^-_j)$.

However, in the case of the GCS (3.15), using standard expressions (like Eqs.
(3.23b)) [11,13] for their overlap integrals, one gets a more complicated
expression
$$
\gamma = \gamma^{(0)} +\gamma^{(1)} + \gamma^{(2)}  \eqno (5.6)$$
where
$$    \gamma^{(0)}=2<P_0> \oint_C  \sin^2\frac{\theta}{2}d\varphi, \quad
2<P_0>= \sum_{j=1}^{m}\left(|\alpha_{j}^{+}|^2-|\alpha_{j}^{-}|^2\right),
\eqno (5.7a)$$
$$    \gamma^{(1)}= - <P_1>\oint_C  [\sin \theta \cos\varphi d\varphi +
\sin \varphi d\theta],\quad
<P_1>= Re [\sum_{j=1}^{m}\alpha_{j}^{-}(\alpha_{j}^{+})^*],
\eqno (5.7b)$$
$$    \gamma^{(2)}= <P_2>\oint_C  [\sin\theta \sin\varphi d\varphi -
\cos\varphi d\theta], \quad
<P_2>= -Im [\sum_{j=1}^{m}\alpha_{j}^{-}(\alpha_{j}^{+})^*]
\eqno (5.7c)$$

As is seen from Eqs. (5.1), (5.6), (5.7), the total geometric phase for
Glauber's GCS is the sum of single ST mode contributions (because of the
additivity of the $P$-quasispin components) that can be used for
measurements of the total geometric phases with the help of single-mode
interferometric schemes [26,27]. Besides, the contribution of $\gamma^{(0)}$
to the total geometric phase (5.6) is just the classical half the solid
angle subtended by the cyclic evolution loop $C$ on the Poincar\'e sphere,
multiplied by the factor $2<P_0>$  characterizing polarization structure of
the field. If $\sum_{j=1}^{m}\alpha_{j}^{-}(\alpha_{j}^{+})^* =0$ (that
occurs, e.g., in the case when for each $j$ either $\alpha_{j}^{+}$ or
$\alpha_{j}^{-}$ equals zero) then $\gamma^{(1)}$and $\gamma^{(2)}$ vanish,
and Eq.(5.7a) represents the total geometric phase. This is valid,
in particular, for the single-mode states (3.16) associated to elliptically
polarized waves and obtained by means of transmissions of coherent
light beams with a definite circular polarization through polarization
rotators. However,  it is not the case for general GCS (3.15) with reference
vectors (3.14) constrained by conditions (3.17); specifically, even in the
case of one ST mode $\gamma^{(a)}, a=1,2$ do not vanish that reflects a
specific correlation of polarization modes after a transmission of beams
(3.14), with $\alpha^{\pm}_j$ being purely real, through "polarization
rotators" (3.2). In general, Eqs. (5.6)-(5.7) describe a structure (nature)
of influences of polarization rotators on initial Glauber's CS in dependence
on their polarization properties since "energetic" multipliers in these
equations are related to expectation values $<P_{\alpha}>$ of different
components of the polarization quasispin $P_{\alpha}$. Furthermore, as it
follows from Eqs. (5.4)-(5.7), pure quantum states of unpolarized light do
not acquire any geometric phases that can be used as an indicator of these
states.\\

\section{ Conclusion}

 So, we have defined and examined in both mathematical and physical aspects
different types of polarization GCS that enabled to give a quasiclassical
description of polarization structure of quantum light beams and to
determine different sorts of squeezing in polarization quantum optics. On
this way a new classification of polarization states of light beams was
given. The polarization GCS of $SU(2)_p$ group obtained above may be also
applied as many-parametric test wave functions to analyze other aspects of
the polarization quasiclassical description. Among these one should mention
applications in "hidden variable" theories [21,37] and studies of other
types of the polarization, biphoton and quadrature uncertainty relations and
squeezing (cf. [32-34]). Specifically, GCS (3.11) manifest [34] a squeezing
for quadrature components of multimode fields determined according to Ref.
[8]. Besides, the $Y$-biphoton squeezed operators $S_Y(\{\gamma_{ij}\})$
represent a particular case of multimode squeezed operators introduced in
[9] whereas the $X$-biphoton squeezed operators $S_X(\{\zeta_{ij}\}$ are
their skew-symmetric analogs. It is also of interest to consider
generalizations of such GCS obtained when replacing operators
$S_P(\{\xi\}),\;S_X(\zeta_{ij}\})$ and $S_Y(\{\gamma_{ij}\})$ by their
"quantum" analogs associated with nonlinear, purely quantum versions of
Hamiltonians (2.10), (3.2) and polynomial Lie algebras [38].

We have also calculated the occurence of geometric phase in different
polarization GCS due to the cyclic evolution in the space of the
angular coordinates $\theta, \varphi$ of the ``classical'' $P$-quasispin
on the Poincar\'e sphere. The explicit expressions of the geometric
phase are shown to depend on the polarization structure of the reference
state vectors characterized by expectation values of the $P$- quasispin
components that may be useful for a practical identification of these
states. The expressions derived can be used in further investigations of
the geometric phases in quantum optics, e.g., by means of using expansions
of arbitrary pure quantum states of light in series of orthonormalized
states $|p, \mu;n, \lambda>$ and Eq. (5.4).

In conclusion we note that, as whole, results obtained together with those of
Part I give a new insight into polarization stucture of multimode quantum
light beams that opens a prospect for both applications and further
investigations, in particular, search of simplest ways to produce new
polarization states of light, studies of interaction of light in these
states with material media (including optically active molecules and atoms
[34,39-41]) and its propagation through different optical devices as well as
the general problem of the description of the quantum light field phase [24].
Notice, however, that in practice it is easier to realize
GCS corresponding to Eqs. (3.12) (for $\kappa=0$) and (3.13) rather than
Eqs. (3.10)-(3.11) because the latter require either parametric oscillator
crystals with highly anisotropic properties [10] or a synchronous use of two
parametric generators with pumping by one laser source [23]. Therefore,
perhaps, for production of $P$-scalar light it is preferable to combine
more simple schemes of $P_{0}$-scalar light generation together with some
interferometric schemes [10]. It is also of interest to consider different
schemes related to cascade processes (cf. [42]).

Furthermore, squeezing properties described by Eqs. (4.3)-(4.6) and their
analogs can be used for designing different experiments related to the
EPR-paradox and entangled states [20-22] and for precise measurements in
spectroscopy of anisotropic media [23]. Specifically, properties given
by  Eqs. (4.5) may be interpreted as an "inversion" of the classical
EPR-paradox [43]. It is also of interest to determine the orbital angular
momentum for new polarization states as a characterization of their spatial
properties (cf. [44]). Besides, the analysis above can be generalized for
arbitrary boson-fermion  systems with any $SU(n), n\geq 2$ internal
symmetries (cf. [1,15,45]).

\section{Acknowledgements}

Preliminary results of the paper were reported at the International Workshops
"Computer simulation in nonlinear optics" (Volga Laser Tour-93, Moscow-Nizhny
Novgorod-Moscow, June-July 1993) and "Squeezed states and uncertainty
relations-III"(Baltimore, August 1993) and were presented at the Research
Conference on Quantum Optics (Davos, September 1994). I am indebted to
the Joint-Stock Company DALUS for sponsoring my paricipation in the Volga
Laser Tour-93 and to the European Scince Foundation for support of
my participation in the Davos Conference. A partial support is acknowledged
from the Committee for High School of Russia, grant No.94-2.7-1097.
The author thanks C.O. Alley, R.~Bhandari, V.P. Bykov, V.L. Derbov, A.V.
Masalov and J. P. Woerdman for fruitfull discussions.


\begin{thebibliography}{26}
\bibitem{1} V.P. Karassiov(1991), J. Sov. Laser Research, {\bf 12} 147;
 ibidem, p. 431
\bibitem{2} V.P. Karassiov(1993), J. Phys., {\bf A26} 4345 ;\\
-----------(1992) Preprint FIAN N 63, Moscow
\bibitem{3} V.P. Karassiov(1993), Kr. Soobshch. Fiz. FIAN, No 5-6 24 [Bull.
Lebedev Phys. Inst., N5 19]
\bibitem{4} J.M. Jauch and F. Rohrlich(1959), {\it Theory of Photons and
Electrons}, ( London: Addison-Wesley)
\bibitem{5} D. Stoler(1970), Phys. Rev., {\bf D1} 3217
\bibitem{6} H.P. Yuen(1976), Phys. Rev., {\bf A13}  2226
\bibitem{7} D.F. Walls(1983), Nature {\bf 306} 141;\\
 J. Opt. Soc. Am.(1987), {\bf B 3}, N 10;
 J. Mod. Opt.(1987), {\bf 34}, No 6/7 [special issues devoted to
  squeezed states]
\bibitem{8} C.K. Hong and L. Mandel(1985), Phys. Rev., {\bf A32}  974
\bibitem{9} X. Ma and W. Rhodes(1990), Phys. Rev., {\bf A41}  4625
\bibitem{10}
V.P.Karassiov(1994), {\it  Proc. 3 Workshop on Squeezed States and
Uncertainty Relations, Baltimore, August 1993}, (NASA Conf. Publication
CP-3270) 65;\\
------------(1994) Phys. Lett. {\bf A190} 387; \\
------------(1994) J. Rus. [Sov.]  Laser Res. {\bf 15} 391;\\
 V. P. Karassiov and M. Cervantes(1994), Rev. Mex. Fis. {\bf 40} 227.
\bibitem{11} J.R.Klauder and B.-S.Skagerstam(1985), {\it  Coherent States.
Applications in Physics and Mathematical Physics}, (Singapore:
World Scientific);\\
A.M.Perelomov(1986), {\it  Generalized Coherent States and
their Applications}, (Springer, Berlin);\\
W.-M. Zhang, D.H. Feng and R. Gilmore(1990), Rev. Mod. Phys., {\bf 62} 867
\bibitem{12} R. Delbourgo(1977), J. Phys.,{\bf A10}  1837
\bibitem{13} J. Perina(1972), {\it Coherence of Light}, (London: Van
Nostrand);\\
--------(1984) {\it Quantum Statistics of Linear and Nonlinear Optical
Phenomena}, ( Dordrecht: D. Reidel )
\bibitem{14} C.C. Gerry and T. Kiefer (1991), J. Phys., {\bf A24} 3513
\bibitem{15} V.P.Karassiov and L.A.Shelepin (1980), Teor. Mat. Fiz.,
{\bf 45} 54 (Russian)
\bibitem{16} M. Berry, J. Mod. Opt. 34 (1987) 1401
\bibitem{17} Y.Aharonov and J.Anandan(1987), Phys. Rev. Lett., {\bf 58} 1593
\bibitem{18} T.F. Jordan(1988), Phys. Rev., {\bf A38} 1590;\\
J. Samuel and R. Bhandari (1988), Phys. Rev. Lett. {\bf 60} 2339
\bibitem{19} V.P. Karassiov, V.L.Derbov and S.I. Vinitsky (1994),
{\it Computer Simulation in Nonlinear Optics}, (SPIE vol. {\bf 2098}) p. 164
\bibitem{20} Y.H. Shih and C.O. Alley(1988), Phys. Rev. Lett., {\bf 61}
  2921;\\ T.E.Kiess, Y.H. Shih, A.V. Sergienko and C.O. Alley (1993),
  Phys. Rev. Lett., {\bf 61} 3893
\bibitem{21} J.F. Clauser and A. Shimony (1978),  Rep.Progr.Phys. {\bf 41},
 1881;\\
D. Greenberger, M. Horne and A. Zeilinger(1993), Phys. Today,{\bf 46} N8,
Pt.1, 22
\bibitem{22}  T.S. Larchuk, R.A. Campos, J.G. Rarity, R.P. Tapster,
E.Jakeman, B.E.A. Saleh and M.C. Teich (1993), Phys. Rev. Lett.,
 {\bf 70} 1603
\bibitem{23} V.P. Karassiov and A.V. Masalov(1993), Optics and Spectr.(USSR),
 {\bf 74} 551
\bibitem{24} A. Voudras(1990), Phys.Rev., {\bf A41},1653
\bibitem{25} V.P. Karassiov(1987), J. Phys., {\bf A20} 5061
\bibitem{26} D.N. Klyshko (1989), Phys. Lett.{\bf A140} 19; ibidem(1992)
{\bf A163} 349
\bibitem{27} R. Bhandari and J. Samuel(1988), Phys. Rev. Lett.,
{\bf 60} 1211;\\
R. Bhandari (1989), Phys. Lett., {\bf A143} 170;\\
R. Bhandari and T. Dasgupta(1990), Phys. Lett., {\bf A143} 170
\bibitem{28} R. Tanas and S. Kielich (1990), J. Mod. Opt. {\bf 37} 1935;\\
R. Tanas and Ts. Gantsog(1991), JINR preprint No E17-91-304
\bibitem{29} D.R. Truax(1985), Phys. Rev., {\bf D31} 1988
\bibitem{30} G.S. Agarwal(1988), J. Opt. Soc. Am., {\bf B5} 1940
\bibitem{31} K. Wodkiewicz and J.H. Eberly (1985),  J. Opt. Soc. Am.,
{\bf B 2}, 458;\\
P.K. Aravind (1986), J. Opt. Soc. Am., {\bf B 3}, 1712
\bibitem{32} C. Aragone, E. Chalbaud and S. Salamo (1976), J. Math Phys.,
{\bf 17} 1963;\\
M.A. Rashid (1978), J. Math Phys., {\bf 19} 1391
\bibitem{33} (1993) M. Kitagawa and M. Ueda, Phys. Rev., {\bf A 47}, 5138;\\
D.A. Trifonov (1994),  Phys. Lett., {\bf A 187} 284
\bibitem{34} V.P. Karassiov and V.I. Puzyrevsky (1989), J. Sov. Laser Res.
 {\bf 10} 229;\\
-----------(1991), Trudy FIAN (Proc. P.N. Lebedev Phys. Inst.), {\bf 211} 161
\bibitem{35} T.H.Chyba, L.J.Wang, L.Mandel, and R.Simon(1988),
Opt. Lett., {\bf 13} 562;\\
R.Bhandari (1988), Phys. Lett. {\bf A133}  1; \\
---------- (1991) Physica {\it B 175}  111;\\
M.Kitano and T.Yabuzaki (1989), Phys. Rev.{\bf A 142} 321
\bibitem{36} S.Pancharatnam (1956), Proc. Indian Acad. Sci.{\bf A 44}, 247;
\bibitem{37} G.S. Agarwal, D. Home, W. Schleich(1992), Phys. Lett.,
 {\bf A170}, 359
\bibitem{38} V.P. Karassiov(1994), J. Phys., {\bf A 27} 153
\bibitem{39}{\it Recent Advances in Biophoton Research and its
Applications} (1992), eds F.A. Popp, K. H. Li and Q. Gu (Singapore:
World Scientific)
\bibitem{40} V.P. Bykov (1993) {\it Radiation of Atoms in a Resonant
Environment} (Singapore: World Scientific)
\bibitem{41} A.C. Tam and W. Happer (1977), Phys. Rev. Lett., {\bf 38} 278
\bibitem{42} P. Grangier, A. Aspect and J. Vigue (1985) Phys. Rev. Lett.,
{\bf 54} 418
\bibitem{43} A. Einstein, B. Podolsky and N. Rosen (1935), Phys. Rev,
{\bf 47} 777
\bibitem{44}  L. Allen, M. Beijersbergen, R.J.C. Spreeuw and J.P. Woerdman
(1992), Phys. Rev., {\bf A 45} 8185
\bibitem{45} V.P. Karassiov(1992), J. Phys., {\bf A25} 393
\end{thebibliography}
\end{document}